\documentclass[conf]{new-aiaa}
\usepackage[utf8]{inputenc}

\usepackage{graphicx}
\usepackage{amsmath}
\usepackage[version=4]{mhchem}
\usepackage{siunitx}
\usepackage{longtable,tabularx}
\usepackage{subcaption}

\usepackage[linesnumbered,ruled,vlined]{algorithm2e}
\usepackage{algpseudocode}
\usepackage{tikz}
\usetikzlibrary{positioning}
\usetikzlibrary{shapes.geometric}
\usepackage{float}

\title{Simulation Based Reward Function Validation for Multi-Agent On Orbit Inspection}

\author{Patrick Quinn\footnote{PhD Student, Department of Aerospace, Physics and Space Sciences, pquinn2019@my.fit.edu, AIAA Student Member}}

\author{Bala Prenith Reddy Gopu\footnote{PhD Student, Department of Aerospace, Physics and Space Sciences, bgopu2023@my.fit.edu}}

\author{George M. Nehma\footnote{PhD Student, Department of Aerospace, Physics and Space Sciences, gnehma2020@my.fit.edu}}

\author{Dr. Madhur Tiwari\footnote{Assistant Professor, Department of Aerospace, Physics and Space Sciences, mtiwari@fit.edu, AIAA Member}}
\affil{Florida Institute of Technology, Melbourne, FL., 32901}

\begin{document}

\maketitle

\begin{abstract}

A proposed method for the control of groups of inspection spacecraft is Multi-Agent Reinforcement Learning (MARL). While MARL has already been employed for this purpose in previous work, the reward functions used focus on reaching a finite set of predetermined inspection points around the target. In this work, we study and develop a generalized reward function for the MARL inspection task informed by the analysis of 3D reconstructions of inspected objects in orbit. Because the reward function is generalized such that any number of images at arbitrary locations may evaluated, we also allow trained agents to have complete control over when images are collected. With this approach, we gather insights into best practices for not only the specific MARL inspection task, but also gain key takeaways informative to the broader inspection task outside of a MARL context.

\end{abstract}



\section{Introduction}

Concerns over the accumulation of space debris and the accompanying implications for the use of space in the future have led to calls for action to be taken to mitigate further debris build-up, as well as the active removal of debris already in orbit \cite{Nomura2024-jb}. Prior to the removal of orbital debris, it is useful for missions to gather data on debris of interest, in order to properly allocate resources for their removal \cite{Yamamoto2025-np}. On these missions, imagery is often taken which can be used to generate 3D reconstructions of the target debris \cite{Navidzadeh2024-wl, Caruso2023-sp}. The success of these missions is supportive of the usefulness and practicality of orbital inspection already, but some improvements may be necessary to make similar missions feasible when aiming to gather information on the large amount of debris in orbit. One of the issues in scaling up the scope of such missions is the amount of time and effort that goes into planning maneuvers for proximity operations with other objects in orbit, which can take teams of human operators multiple days to do manually \cite{Symonds2014-yk}. Because of this, any methods which can be used to automate the control of the inspection spacecraft greatly increases the feasibility of larger missions. Additionally, rather than using one inspection spacecraft, it may be preferred to use multiple, cutting down on the time taken to accomplish the inspection task as well as adding redundancy to the mission \cite{Bernhard2020-uf}. Without a suitable automated control method, adding more spacecraft would only exacerbate the issues with mission planning already seen today. One proposed method for the control of a group of inspection spacecraft is the utilization of Multi-Agent Reinforcement Learning (MARL). While MARL has already been employed for the spacecraft inspection task \cite{Lei2024-hm,Dunlap2024-gq}, and other work has been done using Reinforcement Learning (RL) for the inspection of small bodies such as asteroids with a single spacecraft \cite{Herrmann2024-fd}, the approaches taken so far have focused on learning policies to reach predetermined inspection points. The distribution of such inspection points, as well as the number of inspection points, is left as a design decision. A potential way to improve upon this approach is by defining reward functions which accommodate an arbitrary number of inspection points at arbitrary positions. In order to ensure the effectiveness and enable the refinement of such reward functions, they can be validated by using the trajectories generated by the MARL agents to gather images of the target in a high-fidelity simulation. Those images may be used to create 3D reconstructions of the target \cite{Li2023-ms}, which may then be compared to a reference model for the target to get both an objective and subjective sense for reward function performance. Shaping of the reward function can then be used to encourage behaviors in the spacecraft group needed to make the inspection mission feasible and safe, such as minimal fuel usage, avoidance of collisions, and effective collaboration. By refining the reward function based on the reconstruction results, the reward function may be parameterized such that a specific level of reconstruction quality and data gathering effort tradeoff may be selected ahead of time.
\\
\\
\\
We provide four main contributions in this work:
\begin{itemize}
    \item An open-ended framework for MARL inspection using PPO, detailing an implementation which accommodates images in any location or quantity.
    \item Details on the methods used to test and refine our MARL inspection implementation to aid further research in orbital inspection.
    \item Ideas for future improvements to RL-based inspection.
    \item Takeaways applicable to both the MARL orbital inspection task as well as the broader orbital inspection task.
\end{itemize}


\section{Theory}

In this section, details are provided on the dynamics and RL/MARL methods used in our work. In order to generate 3D reconstructions, the orbital inspection task we focus on requires capturing many images from a variety of different viewpoints around the inspection target. Because of this, we use coordinate frames and dynamics suitable for proximity operations. For our RL environment, we define all states of spacecraft in the LVLH frame of the target. Commonly used for expressing relative orbital dynamics, the LVLH frame is defined as having the $x$-axis pointing radially outward from the origin of the target, the $y$-axis along the orbital track, and the $z$-axis forming a right-handed orthonormal basis. In our work, we use Hill-Clohessy-Wiltshire (HCW) relative dynamics \cite{Clohessy1960-yl} with control input along the axes of the LVLH frame of the inspection target as the governing dynamics. These dynamics are used for their simplicity and alignment with our simulation environment, where we assume inspection targets with circular orbits, and inspection spacecraft with nearly identical orbits to their target. The HCW relative dynamics with control input are given in Eq.~\ref{eq:Clohessy-Wiltshire_Eqn}.

\begin{align}
    \ddot{x}&=3n^2x+2n\dot{y}+u_x/m\notag\\
    \ddot{y}&=-2n\dot{x}+u_y/m\label{eq:Clohessy-Wiltshire_Eqn}\\
    \ddot{z}&=-n^2z+u_z/m\notag\\
    \text{where:}&\notag\\
    n&=\sqrt{\mu/a^3}\notag\\
    m&=100\,\text{kg}\notag\\
    u_x, u_y, u_z &\in \left[-10\,\text{N},10\,\text{N}\right]\notag
\end{align}

In the Equation, $\mu$ represents the standard gravitational parameter and $a$ is the radius of the orbit. Only translational dynamics are considered in our work, so the state vector for an inspection spacecraft $j$ is given as:

\begin{equation}
    \mathbf{x}_j^\text{T}=\left[x_j,y_j,z_j,\dot{x}_j,\dot{y}_j,\dot{z}_j\right]
    \label{eq:stateDef}
\end{equation}

The main motivation for the use of RL in this paper is to learn policies which map information about the state of the environment to actions in order to maximize the long term reward received. Our RL environment evolves over time as a Markov Decision Process (MDP), where an agent takes information about the environment and uses a policy to affect actions. An MDP is defined in discrete time, with a state transition model defined by the dynamics given in Eq.~\ref{eq:Clohessy-Wiltshire_Eqn}, which are integrated over each time step to provide a discrete dynamics model, as well as internal logic to keep track of lighting conditions and images taken. MDPs are defined by a tuple:

\begin{equation}
    (\mathcal{S},\mathcal{A},P,R,\gamma)
    \label{eq:MDPTuple}
\end{equation}

In the tuple, $\mathcal{S}$ represents the set of states of the environment, $\mathcal{A}$ represents the set of actions available to an agent, $P$ represents a probabilistic state transition function $P(s'|s,a)$, $R$ represents a reward function $R(s,a,s')$, and $\gamma$ represents a discount factor for future rewards.

In order to maximize long term reward in RL, we aim to find the parameters for a policy which maximizes the expected future discounted reward for a policy, seen in Eq.~\ref{eq:longTermReward}.

\begin{equation}
    \mathbb{E}^{\pi}\left[\sum_{t=0}^{\infty}{\gamma^tr\left(s_t,a_t\right)}\right]
    \label{eq:longTermReward}
\end{equation}

In this equation, $\pi$ represents the policy used, $\gamma$ represents a discount factor between $0$ and $1$ used to regulate how much the policy focuses on long- or short-term rewards, and $r$ represents the reward as a function of a state $s$ and action $a$.

In order to train policies for our agents, we utilize \textit{Proximal Policy Optimization} (PPO)\cite{Schulman2017-pi}, a policy gradient method. Policy gradient methods work to approximate the policy directly by maximizing a surrogate objective function at each iteration:

\begin{align}
    L_{t}^{CLIP+VF+S}\left(\theta\right)&=\hat{\mathbb{E}}_{t}\left[L_{t}^{CLIP}\left(\theta\right)-c_1L_{t}^{VF}\left(\theta\right)+c_2S\left[\pi_\theta\right]\left(s_t\right)\right]\label{eq:PPOObj}\\
    \text{where:}&\notag\\
    L_t^{CLIP}\left(\theta\right)&=\hat{\mathbb{E}}_t\left[\min\left(r_t\left(\theta\right)\hat{A}_t,\text{clip}\left(r_t\left(\theta\right),1-\epsilon,1+\epsilon\right)\right)\right]\notag\\
    L_t^{VF}\left(\theta\right)&=\left(V_\theta\left(s_t\right)-V_t^{\text{targ}}\right)^2\notag
\end{align}

Where $c_1$ and $c_2$ are coefficients, $\epsilon$ is a hyperparameter, and $S$ is a bonus for entropy, encouraging exploration. $\theta$ represents the parameters of the policy being trained, which in our implementation are the parameters of a neural network. $\hat{A}_t$ is an approximation of the advantage function. The advantage function, $A(s,a)$, is defined as the difference between the $Q$-value function and the optimal value function:

\begin{align}
    A\left(s,a\right)&=Q\left(s,a\right)-V^*\left(s,a\right)\label{eq:Advantage}\\
   \text{where}:&\notag\\
   V^*\left(s\right)&=\underset{a\in\mathcal{A}}{\max}\left[r\left(s,a\right)+\gamma\sum_{s'\in S}{V^*\left(s'\right)p\left(s'|s,a\right)}\right]\notag\\
   Q\left(s,a\right)&=r\left(s,a\right)+\gamma\sum_{s'\in S}{V^*\left(s'\right)p\left(s'|s,a\right)}\notag
\end{align}

The optimal value function, $V^*\left(s\right)$, gives the expected cumulative discounted reward if an optimal policy is followed from the current state, while the Q-value function gives the expected cumulative discounted reward if the optimal policy is followed after taking a given initial action in the current state.

Our MARL implementation employs a Decentralized Training for Decentralized Execution (DTDE) scheme, training each agent in isolation with PPO. This is in contrast to the popular Centralized Training for Centralized Execution (CTDE) approach used in MARL, where a centralized critic with privileged knowledge is used to evaluate individual agent's policies \cite{Amato2024-kc}. Each agent receives observations about the environment and chooses actions to take simultaneously, without explicit coordination. The actions of each other agent are treated as part of the environment's dynamics, so coordination must be learned implicitly.

\section{Methodology}

Our method for tackling the creation and validation of reward functions for the MARL inspection problem builds off the work that others have done in this area \cite{Bernhard2020-uf,Lei2024-hm,Dunlap2024-gq} by taking inspiration from the methods they have used, then implementing our own candidate reward functions designed to improve the flexibility agents have in accomplishing the inspection task at a desired level of quality. These reward functions are designed to include key parameters for the task, relating to fuel \& time efficiency, image quality, and safety. Most of these metrics can also be used as direct measures of performance, with the quality of the images collected being evaluated by their usefulness in making 3D reconstructions of the target. We generate reconstructions by taking images captured with inspection spacecraft and feeding those images into both COLMAP\cite{schoenberger2016mvs,schoenberger2016sfm}, and Instant-NGP \cite{mueller2022instant}, which produce 3D reconstructions of the object(s) imaged. The reconstruction can then be compared quantitatively through various methods\cite{di-Filippo2024-ig}. More details on our implementation, analysis, and generation of results are given below. 

In our RL environment, each spacecraft is given access to its own state and the states of each other inspection spacecraft. Additionally, each spacecraft is given information pertaining to current lighting conditions, enabling a focus on well-lit imagery. The inspection spacecraft are also given information on the last $n$ images taken across all inspection spacecraft. In following with the approach of other work for MARL orbital inspection \cite{Lei2024-hm}, we employ a hierarchical approach to spacecraft control, relying on two separate and independent agents to handle the control and decision making of each inspection spacecraft. These agents, referred to as upper- and lower-level agents, are responsible for high-level decision making and low-level control, respectively. Upper-level agents are responsible for selecting a target point to reach, when to take images, as well as when to end the episode. Lower-level agents are responsible for generating appropriate thrust commands for reaching given target points while minimizing fuel usage and avoiding collisions with other inspection spacecraft and the target. Overall, the aim is to train policies for agents which optimize fuel usage, time taken to reach a certain level of reconstruction quality, and safety.

With these stipulations, we define the observation spaces for both upper- and lower-level agents, their respective action spaces, and our candidate reward function for testing. We will start by defining the observation space for each lower-level agent. The lower-level observation space is seen in Eq.~\ref{eq:lowerLevelObservations}.

\begin{equation}
    \begin{aligned}
        \mathcal{O}_\text{l}=&\langle (x_{\text{target}},y_{\text{target}},z_{\text{target}}), (x_1,y_1,z_1,\dot{x}_{1},\dot{y}_{1},\dot{z}_{1}), 
        (x_2,y_2,z_2,\dot{x}_{2},\dot{y}_{2},\dot{z}_{2}), (x_3,y_3,z_3,\dot{x}_{3},\dot{y}_{3},\dot{z}_{3})\rangle
    \end{aligned}
    \label{eq:lowerLevelObservations}
\end{equation}

Each lower-level agent is provided the target position determined by its upper-level agent, its own states, as well as the states of each other inspection spacecraft. Next, we define the upper-level agent's observation space in Eq.~\ref{eq:upperLevelObservations}.

\begin{equation}
    \begin{aligned}
        \mathcal{O}_\text{u}=&\langle (x_1,y_1,z_1,\dot{x}_{1},\dot{y}_{1},\dot{z}_{1}), 
        (x_2,y_2,z_2,\dot{x}_{2},\dot{y}_{2},\dot{z}_{2}), (x_3,y_3,z_3,\dot{x}_{3},\dot{y}_{3},\dot{z}_{3}),\hat{r}_{\text{self}},\hat{r}_{\text{Sun}},c_{\text{Earth shadow}},i_{\text{images taken}},\ldots\\&(x_{\text{near image 1}},y_{\text{near image 1}},z_{\text{near image 1}}),\ldots,(x_{\text{near image 3}},y_{\text{near image 3}},z_{\text{near image 3}}),\ldots\\&(x_{\text{image 1}},y_{\text{image 1}},z_{\text{image 1}}),\ldots,(x_{\text{image n}},y_{\text{image n}},z_{\text{image n}})\rangle
    \end{aligned}
    \label{eq:upperLevelObservations}
\end{equation}

Each upper-level agent is provided with access to the states of all inspection spacecraft, the unit vector from the target towards itself, the Sun's unit vector from the target, whether the target is currently in Earth's shadow, the total number of pictures taken, the positions of the 3 nearest images taken across all agents, and the last $n$ images taken. For our tests, we set $n$ to $100$ images, which was observed to allow for a good reconstruction if the images taken were of sufficient quality. In order to keep a consistent observation size when there are less than $n$ pictures taken, the positions of pictures not yet taken are set to $(0,0,0)$.

The action space of the lower-level agent, $\mathcal{A}_\text{l}$, is defined such that it controls the thrust output for the spacecraft along each axis of the LVLH reference frame. It is defined in Eq.~\ref{eq:lowerLevelActions}. 

\begin{equation}
	\mathcal{A_{\text{l}}}=\left\langle u_x,u_y,u_z \right\rangle
	\label{eq:lowerLevelActions}
\end{equation}
 
The action space of the upper-level agent, $\mathcal{A}_\text{u}$, is defined such that it outputs a target point for the lower-level controller to navigate to, a signal to capture an image, and a signal to end the episode. It is defined in Eq.~\ref{eq:upperLevelActions}.

\begin{equation}
	\mathcal{A_{\text{u}}}=\left\langle x_{\text{target}},y_{\text{target}},z_{\text{target}},b_{\text{capture}},b_{\text{end episode}}\right\rangle
	\label{eq:upperLevelActions}
\end{equation}
 
Because the rest of the action space is continuous, the discrete nature of the signals to capture images and to end the episode is handled by triggering those events when the value given for them rises above a certain threshold. While methods for hybrid continuous-discrete action spaces have been developed in recent work \cite{Neunert2020-lc,Li2021-vh}, we do not implement these as it is believed that the definitive and rare nature of these decisions will make the threshold method chosen sufficient.

With the observations defined, we define the reward function arrived at to perform our tests. Work already done in MARL orbital inspection places inspection points roughly in a sphere \cite{Lei2024-hm,Dunlap2024-gq}, so we designed reward functions which will encourage similar behavior, but with greater flexibility. Lighting is an important consideration when taking pictures in orbit, so we also factor it into the reward function. Because of its relevance to both upper- and lower-level agent rewards, we first define the reward component related to image value. During each run of the RL environment, the total image value is updated at each time step, and agents then received rewards based on the difference in image value from the last time step. In Eq.~\ref{eq:AgentImageValue}, the image value attributed to each inspection spacecraft is defined. 

\begin{align}
    \mathcal{V}_{\text{spacecraft $j$ images}}&=\sum_{i\in\mathcal{I}_j}{S_{\text{distance}}^{(i)}\cdot S_{\text{separation}}^{(i)}\cdot S_{\text{lighting}}^{(i)}\cdot S_{\text{self shadow}}^{(i)}}\label{eq:AgentImageValue}\\
    \text{where:}&\notag\\
    \mathcal{I}_j&\ \text{is the set of image indices attributed to spacecraft }j\notag
\end{align}


In Eq.~\ref{eq:AgentImageValue}, $\mathcal{V}$ represents the value associated with the set of images taken by spacraft $j$, and $S^{(i)}$ represents the score of a specific metric of an image with index $i$. The rationale behind the definition of each image's value, and by extension, the value of sets of images, either from a specific spacecraft or across all spacecraft, revolved around creating scoring functions which were very close to $1$ under ideal image conditions, then quickly approached $0$ otherwise. By multiplying individual scoring functions together, it was possible to only reward images which met all the criteria set. We define the scoring for the distance an image was taken at as:

\begin{equation}
    S_{\text{distance}}^{(i)} = e^{-w_{\text{distance sensitivity}}\cdot\left(\left\|\vec{r}_{\text{im $i$}}\right\|_2-c_{\text{targeted distance}}\right)^4}
    \label{eq:distance}
\end{equation}

By selecting values for the distance sensitivity ($w_{\text{distance sensitivity}}$) and the targeted distance ($c_{\text{targeted distance}}$), the range of acceptable image distances and the average accepted image distance can be selected. The scoring metric for image separation, designed to only give rewards for images which are taken from different angles to the target as any other image, is defined next:

\begin{align}
    S_{\text{separation}}^{(i)} &= 1-e^{-\left(w_{\text{im separation}}\cdot\theta_{\text{closest im}}^{\left(i\right)}\right)}\label{eq:separation}\\
    \text{where:}\notag\\
    \theta_{\text{closest im}}^{\left(i\right)} &=\underset{i\neq k}{\min}\arccos\left(\hat{r}_{\text{im }i}\cdot\hat{r}_{\text{im }k}\right),\quad k \in\bigcup_{j=1}^{3}\mathcal{I}_j\notag
\end{align}

The scoring function for image separation, $S_{\text{separation}}^{(i)}$, along with the fact that rewards for imagery come from the change in total image value from the last time step, means that agents can receive a penalty for taking images too similar to other images. By taking an image too close to another image, the values of both will drop, which can result in a penalty being applied. While not quite accurate to the inspection task, where redundant images can be deleted, we maintained this aspect of the reward function to further encourage good separation between images. 

Finally, the metrics to score lighting are designed to measure the effectiveness of angle the image was taken at with respect to the sun. Images taken with a small angle relative to the direction the Sun receive higher scores for taking an image of the well lit side of the inspection target, with the caveat that images taken in exact alignment with the sun receive no reward as the inspection spacecraft would be casting a shadow on the inspection target. The equations are given below:

\begin{align}
    S_{\text{lighting}}^{(i)} &= 0.1\cdot b_{\text{Earth's shadow}}^{(i)}\cdot\cos\left(\theta_{\text{Sun}}^{(i)\, 2}/\pi+9\right)\label{eq:lighting}\\
    S_{\text{self shadow}}^{(i)} &= 1-e^{-\theta_{\text{Sun}}^{(i)}/w_{\text{self shadow}}}\label{eq:selfShadow}\\
    \text{where:}&\notag\\
        \theta_{\text{Sun}}^{(i)} &=\arccos\left(\hat{r}_{\text{im}}^{(i)}\cdot\hat{r}_{\text{Sun}}^{(i)}\right)\notag\\
    b_{\text{Earth shadow}}^{(i)} &=
    \begin{cases}
        0\ \text{if image $i$ was taken while in Earth's shadow}\\1\ \text{otherwise}
    \end{cases}\notag
\end{align}

We also define the total value of images taken by all inspection spacecraft in Eq.~\ref{eq:TotalImageValue}. A table of the constants and weights used in our implementation of the image value equations is given in Table~\ref{tab:imageValueConstants}.

\begin{align}
    \mathcal{V}_{\text{all images}} &=\sum_{j=1}^{3}{\mathcal{V}_{\text{spacecraft $j$ images}}}\label{eq:TotalImageValue}
\end{align}

\begin{table}[H]
	\centering
	\caption{\label{tab:imageValueConstants} Image Value Constants}
	\begin{tabular}{|l|c|}
		\hline
		\textbf{Constant}&\textbf{Value}\\
		\hline
		$w_{\text{distance sensitivity}}$&$1\times 10^{-5}$\\
		\hline
		$c_{\text{targeted distance}}$&$100\,\text{m}$\\
		\hline
		$w_{\text{im separation}}$&$6$\\
		\hline
		$w_{\text{self shadow}}$&$0.03$\\
		\hline
	\end{tabular}
\end{table}

Plots of the image value components and their relevant parameters are provided below in Fig.~\ref{fig:RewardFunctions}. Because $S_{\text{lighting}}^{(i)}$ and $S_{\text{self shadow}}^{(i)}$ share parameters, we multiply the scores together as they are in Eq.~\ref{eq:AgentImageValue}. 

\begin{figure}[H]
	\centering
	\includegraphics[width=1.0\textwidth]{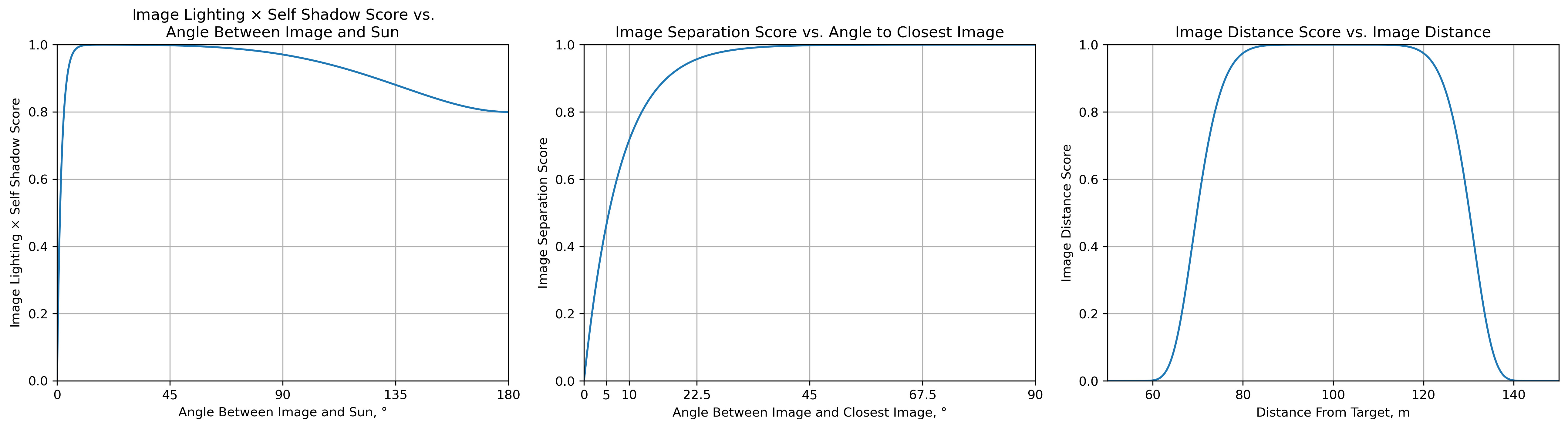}
	\caption{Image value components plotted against their relevant parameters.}
	\label{fig:RewardFunctions}
\end{figure}

Each agent's reward for imagery is given as a mixture of the difference in the total team image value and the difference of that individual agent's image value from the last time step. This is formalized in Eq.~\ref{eq:spacecraftImageReward}. The mixing of team and individual rewards proved essential in discouraging laziness while still encouraging effective teamwork.

\begin{align}
	r_{\text{spacecraft }j\text{ images}} &=c_{\text{mixing coeff}}\cdot\Delta\mathcal{V}_{\text{all images}}+ \left(1-c_{\text{mixing coeff}}\right)\cdot\Delta\mathcal{V}_{\text{spacecraft $j$ images}}\label{eq:spacecraftImageReward}\\
	\text{where:}\notag\\
	\Delta\mathcal{V}_{\text{all images}} &=\mathcal{V}_{\text{all images}}^{\left(t\right)}-\mathcal{V}_{\text{all images}}^{\left(t-1\right)}\notag\\
	\Delta\mathcal{V}_{\text{spacecraft $j$ images}} &=\mathcal{V}_{\text{spacecraft $j$ images}}^{\left(t\right)}-\mathcal{V}_{\text{spacecraft $j$ images}}^{\left(t-1\right)}\notag\\
	c_{\text{mixing coeff}} &= 0.5\notag
\end{align}

With the reward component relating to image value defined, the rest of the reward functions used for upper- and lower-level agents are now able to be defined. First, for the upper-level agents, the reward is given in Eq.~\ref{eq:UpperLevelReward}. A table of the relevant weighting for each component can be found in Table~\ref{tab:upperRewardWeights}.

\begin{equation}
	r_{\text{u, spacecraft }j}=w_{\text{images}}\cdot r_{\text{spacecraft }j\text{ images}}-w_{\text{time}}\cdot \left(t-t_0\right)-w_{\text{fuel}}\cdot\Delta t\cdot\left\|\mathbf{u}_{\text{spacecraft }j}^{\left(t-1\right)}\right\|
	\label{eq:UpperLevelReward}
\end{equation}

\begin{table}[h]
	\centering
	\caption{\label{tab:upperRewardWeights} Upper-level agent reward weights}
	\begin{tabular}{|l|c|}
		\hline
		\textbf{Weight}&\textbf{Value}\\
		\hline
		$w_{\text{images}}$&$100$\\
		\hline
		$w_{\text{time}}$&$5\times10^{-4}$\\
		\hline
		$w_{\text{fuel}}$&$4$\\
		\hline
	\end{tabular}
\end{table}

Upper-level agents are given rewards/penalties for how the value of team and individual images changes, an increasing penalty as elapsed time increases, and penalties for the fuel used in the last time step by the relevant lower-level agent. Fuel usage is penalized in upper-level agents to discourage them from asking lower-level agents to move to distant points too quickly, which would directly go against the lower-level agent's objectives to minimize fuel usage. This was done to help synchronize the objectives and interaction between the upper- and lower-level agents. The time penalty used was kept very low for our tests because we chose to focus on having agents stop the inspection task once the diminishing returns in image value were outweighed by fuel usage penalties. If image quality wasn't as important, however, or there were time constraints on the inspection mission, this value could be raised to be more significant.

Lower-level agents, responsible for moving the spacecraft to the points designated by the upper-level agents while remaining safe and fuel efficient, are given the reward function seen in Eq.~\ref{eq:LowerLevelReward}. A table of the values used for our training is provided in Table~\ref{tab:lowerRewardWeights}.

\begin{align}
	r_{\text{l, spacecraft }j}&=-w_{\text{target point error}}\cdot c_{\text{target point error}}-w_{\text{fuel}}\cdot\Delta t\cdot\left\|\mathbf{u}_{\text{spacecraft }j}^{\left(t-1\right)}\right\|-\ldots\label{eq:LowerLevelReward}\\
	&w_{\text{collision}}\cdot b_{\text{spacecraft }j\text{ collision}}-w_{\text{too far}}\cdot b_{\text{too far}}+w_{\text{images}}\cdot r_{\text{spacecraft }j\text{ images}}\notag\\
	\text{where:}\notag\\
	c_{\text{target point error}}&=\left\|\mathbf{x}_{\text{target}}^{(t-1)}-\mathbf{x}_{j}^{(t)}\right\|\notag\\
	b_{\text{spacecraft }j\text{ collision}}&=\begin{cases}1\text{ if spacecraft }j\text{ is within }25\,\text{m of another spacecraft or the inspection target}\\0\text{ otherwise}\end{cases}\notag\\
	b_{\text{too far}}&=\begin{cases}1\text{ if }\left\|\mathbf{x}_{j}^{(t)}\right\|\geq275\,\text{m}\\0\text{ otherwise}\end{cases}\notag
\end{align}

\begin{table}[H]
	\centering
	\caption{\label{tab:lowerRewardWeights} Lower-level agent reward weights}
	\begin{tabular}{|l|c|}
	\hline
	\textbf{Weight}&\textbf{Value}\\
	\hline
	$w_{\text{target point error}}$&$1$\\
	\hline
	$w_{\text{fuel}}$&$4$\\
	\hline
	$w_{\text{collision}}$&$5000$\\
	\hline
	$w_{\text{too far}}$&$500$\\
	\hline
	$w_{\text{images}}$&$100$\\
	\hline
\end{tabular}
\end{table}

This lower-level reward function gave penalties for distance to the last given target point, fuel usage, near collisions, moving too far from the inspection target, and gave rewards for increasing image value. Image value rewards, while not directly under the control of the lower-level agents, are important to encourage behaviors in the lower-level agents which would enable the upper-level agents to take effective images. This served as a method to improve alignment between upper- and lower-level agent goals.

Due to the significant number of behaviors that result in penalties, and the selective nature of how image rewards are given, training effective policies in the environment proved to be difficult. Leaving the region of operations so the episode terminated quickly or not taking any actions at all were both very common policies. In order to train past these policies, it is necessary to adopt a curriculum learning \cite{Bengio2009-uf} inspired approach. Our implementation revolved around easing the penalties that agents would otherwise incur during early training. To accomplish this, fuel usage started with near $0$ weighting and flat rewards were given at every time step to encourage staying within the region of operations. The weighting for the fuel penalties could gradually be increased as the agents converged on suitable policies for each parameter set, up to the final values listed in Tables~\ref{tab:upperRewardWeights} and \ref{tab:lowerRewardWeights}. The process of increasing fuel penalties had to be done slowly, as increasing fuel penalties too quickly resulted in agents reverting to policies where no fuel was expended. In addition to the gradual increase in fuel penalties, flat rewards for staying within the region of operations were able to be quickly removed once agents learned to stay within the region.

\section{Simulation Details}

The definitions of dynamics, observation \& action spaces, rewards, and all the background logic needed for them informed the creation of a gym environment \cite{Towers2025-yu} for use with Ray RLLib \cite{liang2018rllib,liang2021rllib}, which handled simulation of the environment and training of policies. The specific scenario for this work used a $300\,\text{km}$ orbital altitude above Earth. We consider each spacecraft as having a mass of $100\,\text{kg}$ and a maximum thrust along each axis of the LVLH frame of $10\,\text{N}$, with no consideration for mass change due to fuel use. Within the RL environment, true anomaly of the target is initialized randomly along with the relative positions of each inspection spacecraft within the $300\,\text{m}$ radius region of operations. Initialized positions for the inspection spacecraft are constrained to be further than $25\,\text{m}$ from any other inspection spacecraft or the inspection target to avoid initial unavoidable near collisions. All logic within the gym environment is updated in intervals of $10\,\text{s}$, with dynamics propagated using the fourth-order Runge-Kutta (RK4) method. Alongside the relative dynamics, the orbital dynamics of the inspection target were propagated in order to calculate lighting conditions.

\begin{figure}[h!]
	\centering
	\includegraphics[width=1.0\textwidth]{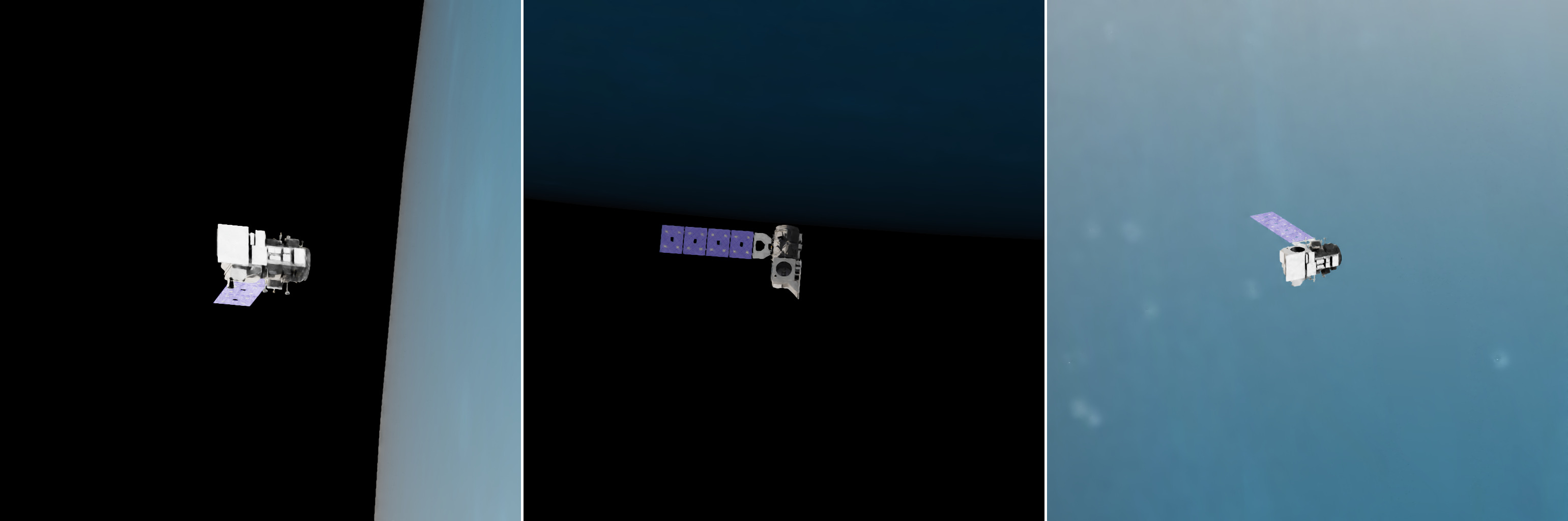}
	\caption{Sample images produced by inspection spacecraft within Isaac Sim.}
	\label{fig:SimulationPhotos}
\end{figure}

Once effective policies are trained, the information from RL episodes pertaining to the trajectories, control effort, image locations, image attributions, and lighting conditions is saved to python data files. These files are used in a script for NVIDIA Isaac Sim \cite{NVIDIA_Isaac_Sim}, in order to replicate the episode in a photo-realistic simulation environment. Our implementation of the environment in Isaac Sim transferred all coordinates to the ECI frame, then placed models of Earth, inspection target \cite{Sanders2025-xb}, and a distant light source representing the Sun as they were recorded in the RL episode being replicated. A camera was then moved to each location an image was taken and captured an image looking at the centroid of the inspection target. As the camera moved, the position of the inspection target in its orbit, and Earth's rotation were updated in accordance with the information from the RL episode. As an example of what the simulation produced, the latest images taken by each inspection spacecraft in the simulation at a sampled point in time can be seen in Fig.~\ref{fig:SimulationPhotos}.

Once images were captured in Isaac Sim, the final step in our analysis was to use software capable of generating 3D reconstructions from images. The two pieces of software used in our analysis were COLMAP, which uses photogrammetry techniques to estimate camera poses and generate reconstructions, and Instant-NGP, which takes given camera poses and uses neural networks to predict density and color along rays (in this case density is a measure of whether an object is present in a certain location). We wrote the file containing the camera positions for Instant-NGP manually with the same python data files used to replicate episodes and take images in Isaac Sim. The predictions from Instant-NGP can be used to generate novel views of the object imaged, or can be used to generate 3D reconstructions. These reconstructions can then be analyzed qualitatively and quantitatively to determine the effectiveness of trained policies for inspection. Because of the motion of the inspection target and the rotation of Earth, the scene behind the spacecraft was inconsistent, which typically presents a challenge for reconstruction software. In order to improve the ability to get reconstructions from the software used, we filtered out all parts of the images aside from the inspection target, setting the irrelevant pixels as transparent. This is trivially done in Isaac Sim, where a limit can be placed on the distance a camera can "see", but it would also be relatively simple to implement in a real-world deployment, with the use of ML-based image segmentation models such as the Segment Anything Model \cite{kirillov2023segany}, or with a variety of distance sensors coupled with image collection.

\section{Results}

\begin{figure}[H]
	\centering
	\includegraphics[width=0.4\textwidth]{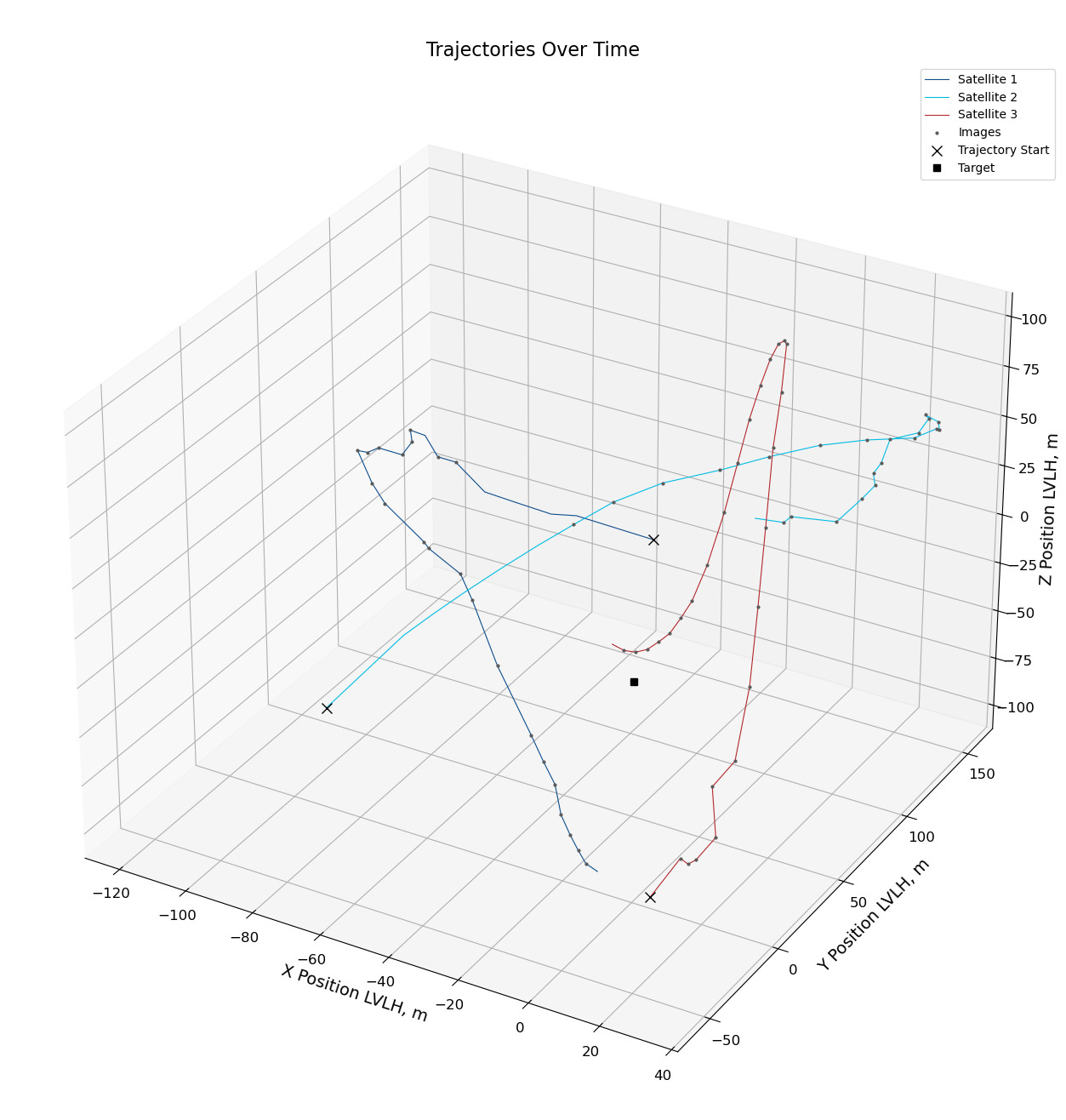}
	\caption{Sample trajectories taken by inspection spacecraft.}
	\label{fig:3DTrajectory}
\end{figure}

\begin{figure}[H]
	\centering
	\includegraphics[width=1.0\textwidth]{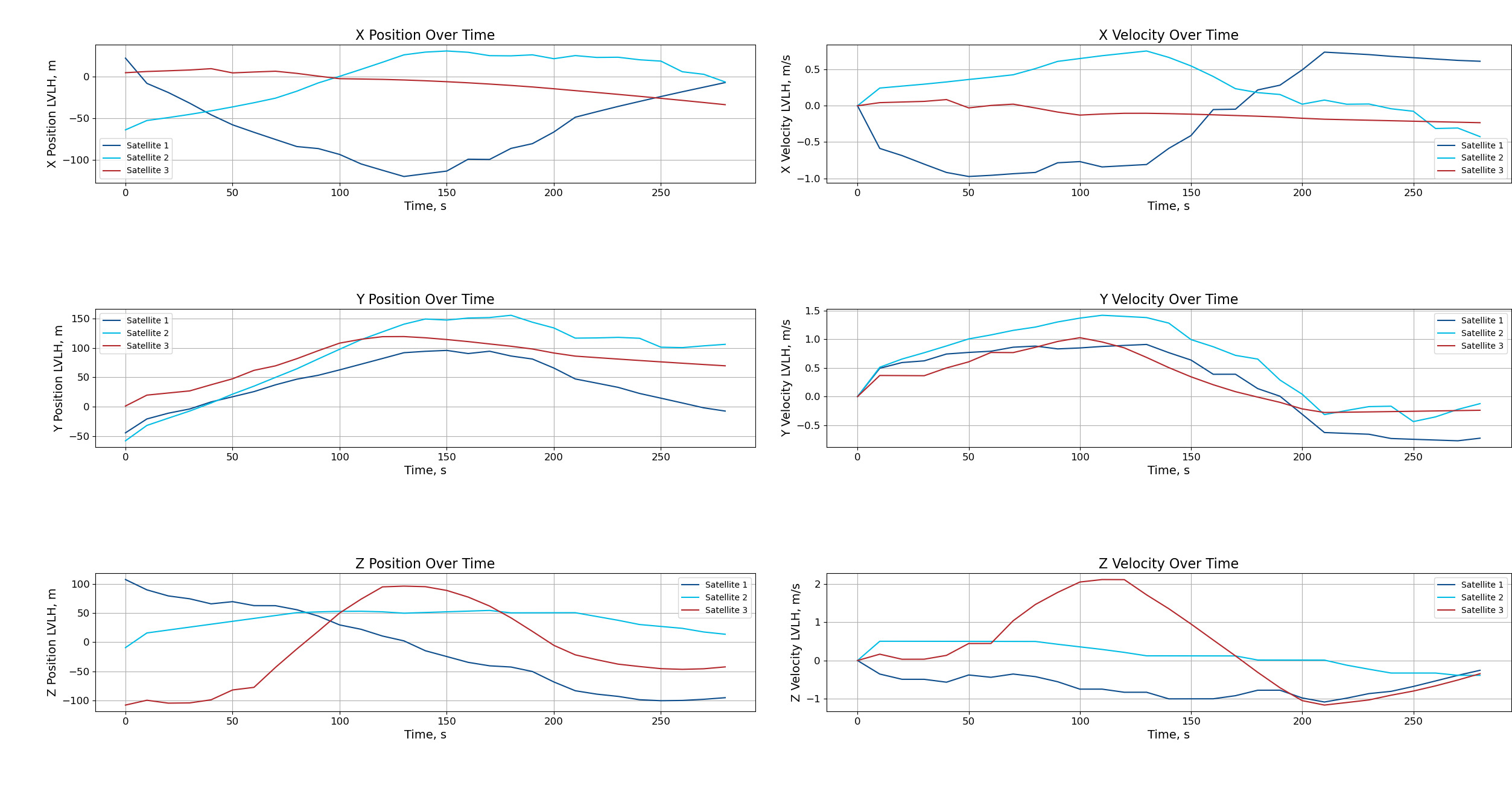}
	\includegraphics[width=1.0\textwidth]{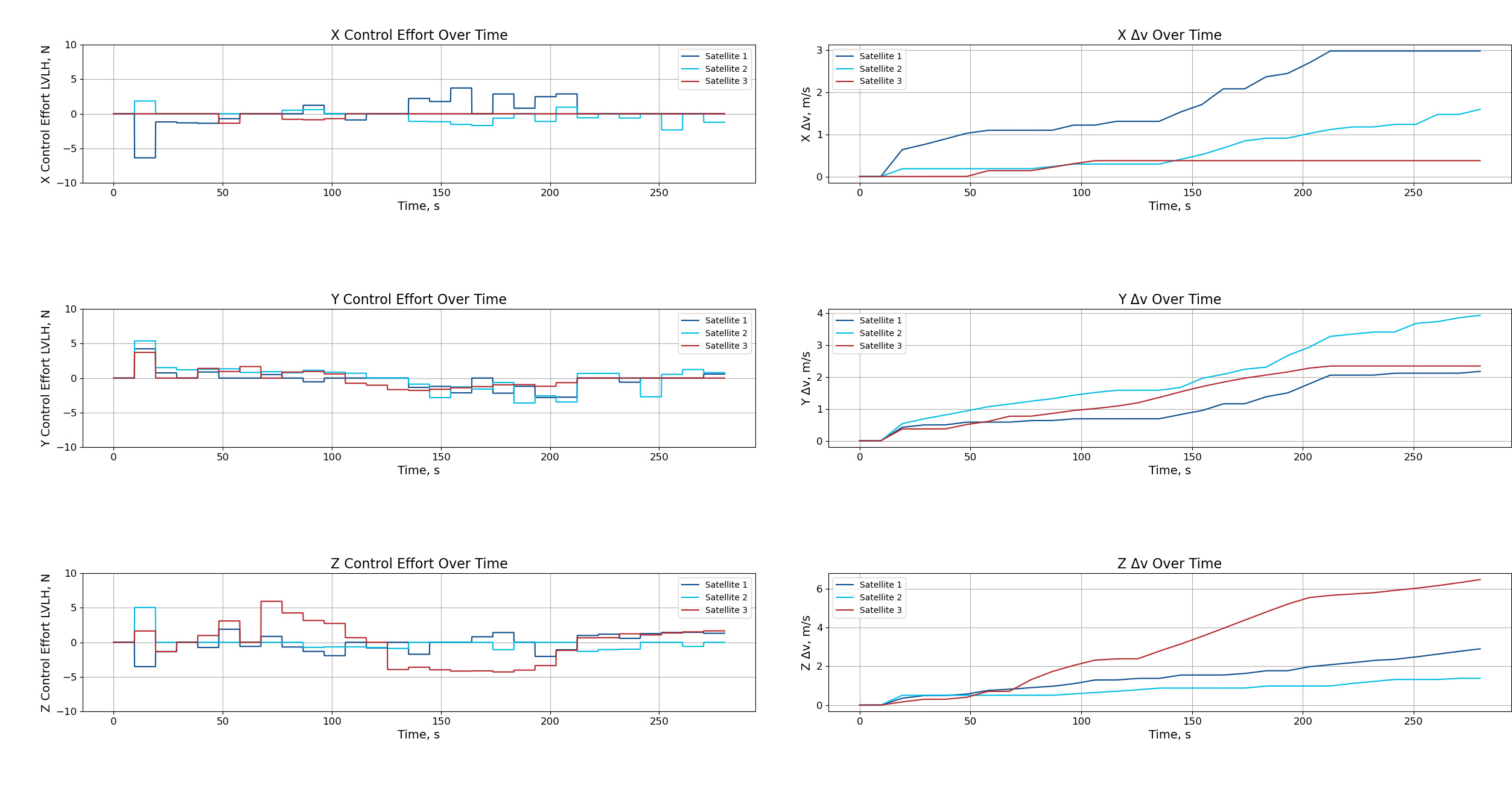}
	\caption{State and control histories of inspection spacecraft.}
	\label{fig:stateControlHist}
\end{figure}

In this section, we cover observations on general behaviors, along with the benefits and shortcomings of our approach to MARL inspection. First, we will discuss how changes in reward function constants and weights affected the general behaviors of policies generated. The most impactful parameter, the weight for the fuel usage penalty, had to be adjusted carefully in order to maintain reasonable policies. At low values, the trajectories formed by the inspection spacecraft appeared to be random, crossing back over areas they had already been countless times in what appeared to be a completely arbitrary manner. Increasing the weight of fuel penalties ($w_{\text{fuel}}$) from just $3$ to $4$ was enough to influence the policy to produce trajectories which were much more efficient. With a fuel penalty weight of $4$, inspection spacecraft still rely on active control for their trajectories, but their thrust outputs focused on oscillations in the LVLH $z$-axis while utilizing the natural HCW dynamics to assist with procession in the $xy$-plane. An example of such a trajectory, along with relevant plots for the state and control histories can be found in Figs. \ref{fig:3DTrajectory} and \ref{fig:stateControlHist}, respectively. Increasing the penalty for fuel usage past this point quickly led to policies which solely rely on natural dynamics for movement between image locations. Using $40\%$ of a perfect image's value as a penalty per $1\,\text{m/s of }\Delta v$ used resulted in the actively controlled inspection trajectories seen in the Figure, while $50\%$ of a perfect image's value seemed to be the upper limit on fuel penalty before agents drifted with the natural dynamics. In our testing, we also found that attempts to increase the desired image separation past the parameters used in Table~\ref{tab:imageValueConstants} discouraged image taking to the point that agents stopped taking images altogether. Agents developed strong preferences for taking images on the side of the inspection target facing the sun, and the image value function had to be tweaked to give more value to images taken on the poorly lit side of the inspection target to what is seen in Eq.~\ref{eq:AgentImageValue}. From real world experience \cite{Yamamoto2025-np,Navidzadeh2024-wl}, as well as what we observed in our simulated environment, images taken on the shadowed side of an inspection target are often still workable for inspection purposes, even if they aren't optimal.

\begin{figure}[H]
	\centering
	\includegraphics[width=.5\textwidth]{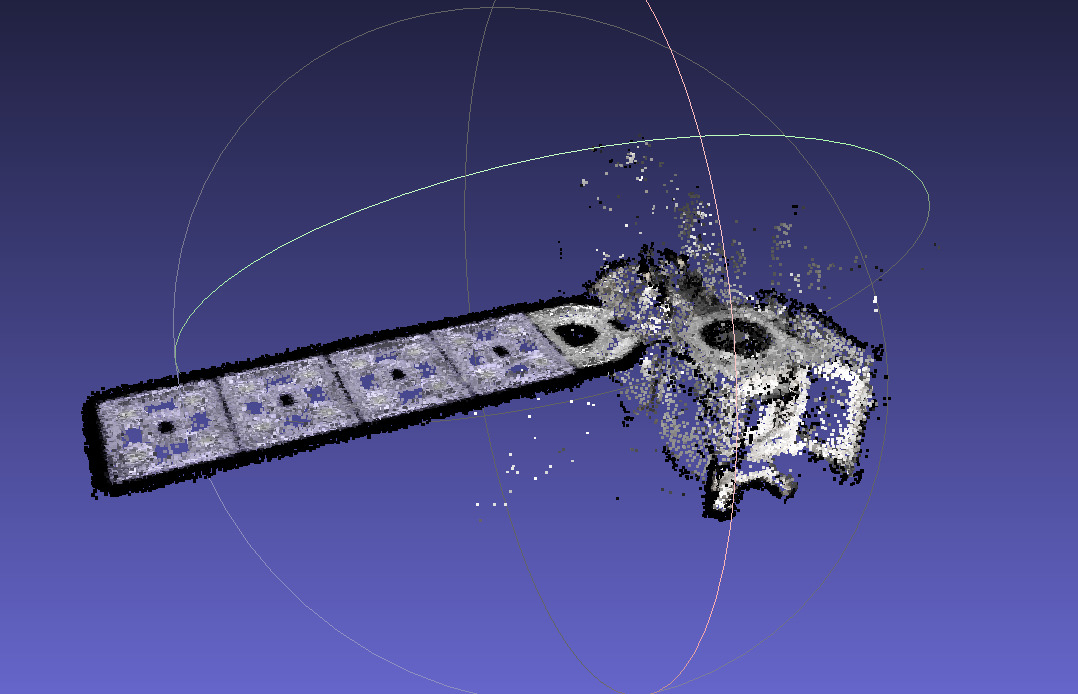}
	\caption{Model reconstruction produced by COLMAP.}
	\label{fig:COLMAPReconstruction}
\end{figure}

\begin{figure}[H]
	\centering
	\begin{subfigure}[b]{0.45\textwidth}
		\centering
		\includegraphics[width=\textwidth]{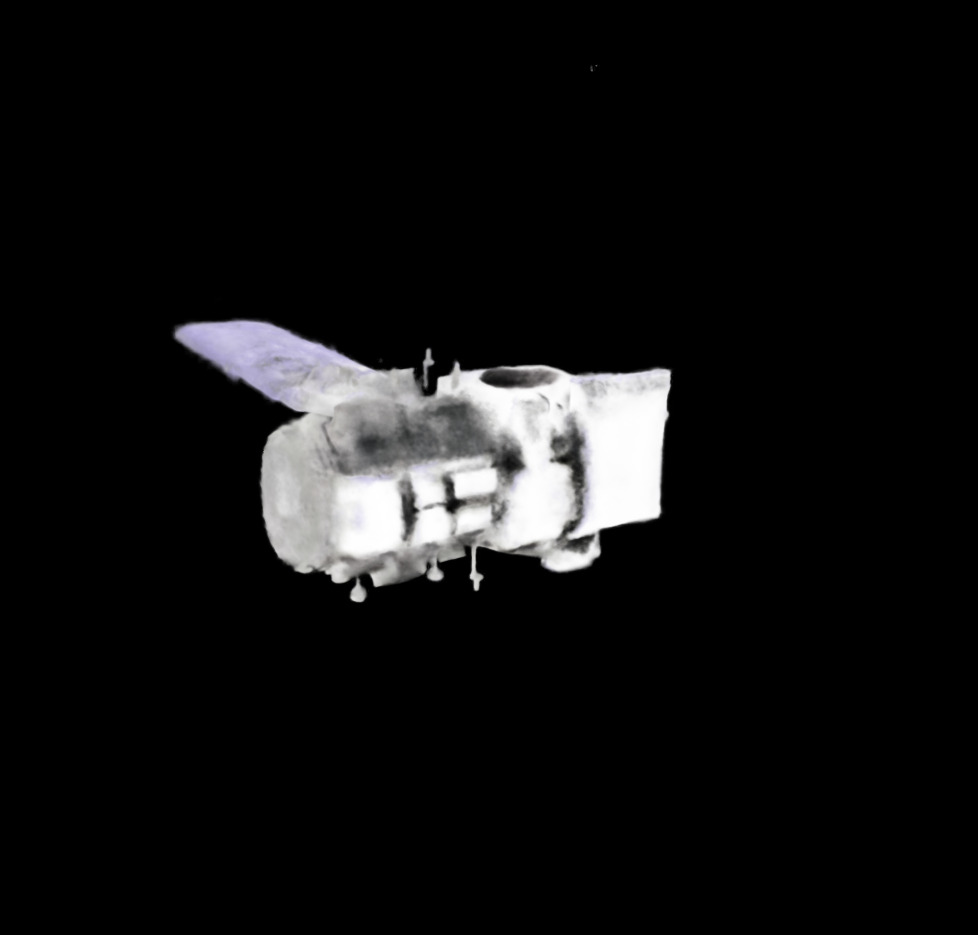}
	\end{subfigure}
	\hfill 
	\begin{subfigure}[b]{0.45\textwidth}
		\centering
		\includegraphics[width=\textwidth]{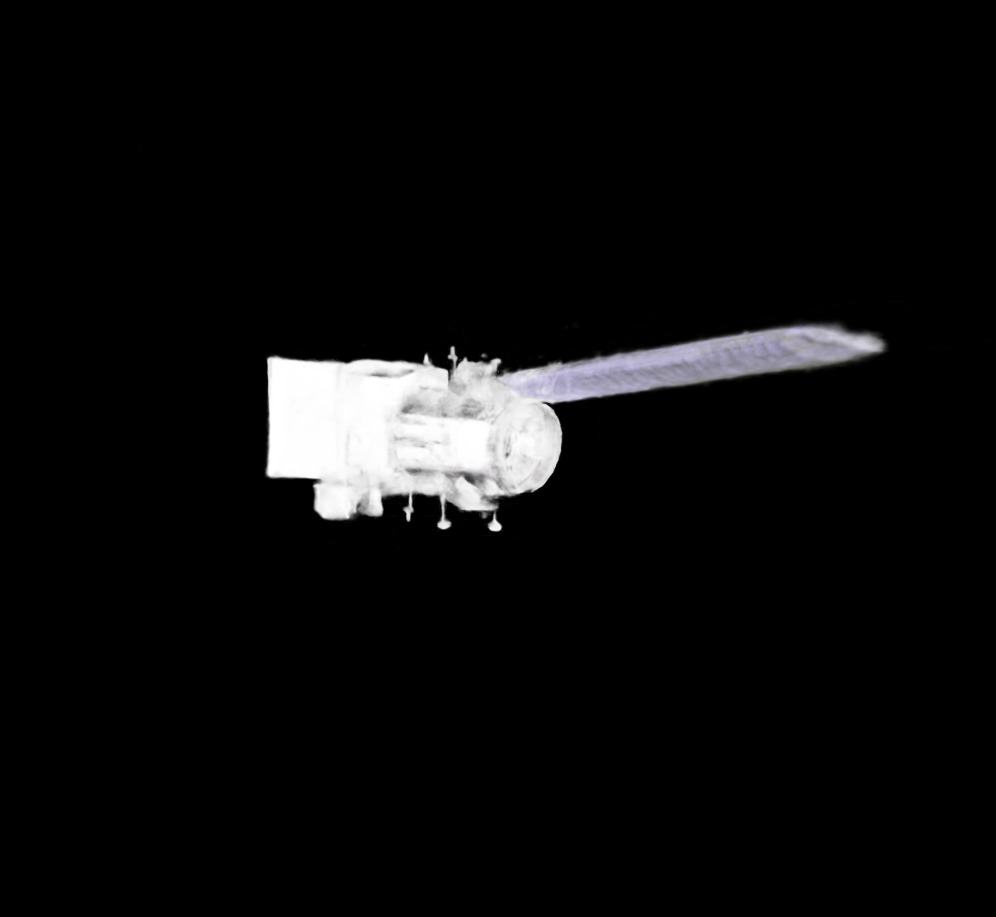}
	\end{subfigure}
	\caption{Model reconstruction produced by Instant-NGP.}
	\label{fig:INGP}
\end{figure}

After generating inspection images and feeding them into COLMAP and Instant-NGP, we observed the following results. First, while COLMAP provided a clean reconstruction of some sections of the inspection target, many sections that were imaged went un-reconstructed. We suspect this may be due to changes in lighting conditions as the inspection target orbited Earth making the photogrammetry more difficult. A sample reconstruction from COLMAP is seen in Fig.~\ref{fig:COLMAPReconstruction}. Using the same dataset, Instant-NGP produced a fuzzier but more complete reconstruction of the target. This can be seen in Fig.~\ref{fig:INGP}. While harder to see in the Instant-NGP results, there is still a portion of the inspection target's bus which wasn't reconstructed. The results seen here were promising, and it is worth noting that a model such as Neuralangelo \cite{Li2023-ms} could produce higher fidelity reconstructions at the cost of an increase in computation time.

Generally, the results seen are promising, however the trajectories generated still have some deficiencies. The difficulty in training decent policies suggests the possibility that some combination of parameters and curriculum learning methodology that went untested may produce better results than these, and we believe that such experiments are worth pursuing. Additionally, there are some improvements which may improve the efficacy of training. One possible improvement would be the definition of angular sectors around the target, then giving the number of images in each sector as an observation rather than the exact position of each image. While losing some data, this would make learned policies significantly more generalizable and robust to unseen image configurations. We tested such an improvement and couldn't get a good policy from it, but given the difficulties in training, we believe this is an approach still worth pursuing. Another possible improvement would be using a more traditional control method such as MPC for low-level control and collision avoidance, discarding the lower-level agents entirely. This would eliminate the complex interplay between upper- and lower-level agent policies, potentially simplifying training while also improving the optimality and robustness of control. The final improvement which could help if implemented is CTDE, but the extent to which this would improve training and results is hard to say.

\section{Conclusion}

The MARL methods used in this work generated policies and results which support this being an effective method of control for inspection of objects on orbit, but it is also clear that additional refinement is needed of the techniques presented here to get the type of results and policies that would warrant deployment in space. We do believe, however, that the use of MARL offers value to the task solely due to the ability it gives mission designers to explore general lessons and strategies. The takeaways from this work could be applied to the inspection task with any number of traditional controllers, and done so in a way that the behaviors were optimized and safe.

Future work in this direction which is of particular importance is in RL safety, or the development of assurances for the safety of RL systems. Developments here are crucial to the applicability of RL in close proximity operations in space, as accidents are incredibly costly and dangerous to other spacecraft. Further future work, which may prove computationally infeasible, would be full automation of the image gathering, reconstruction, and reconstruction analysis processes in order to provide a direct RL reward for reconstruction quality. For the resources available to us, this wouldn't have been a feasible approach to the RL problem, but as methods of reconstruction improve the feasibility of such an approach will increase.


\bibliography{bibliography}

\end{document}